# Dragon-Kings, Black Swans and the Prediction of Crises

Didier Sornette [a,*]

[a] *ETH Zurich*
*Department of Management, Technology and Economics*
*Kreuzplatz 5, CH-8032 Zurich, Switzerland*

**Abstract**
We develop the concept of "dragon-kings" corresponding to meaningful outliers, which are found to coexist with power laws in the distributions of event sizes under a broad range of conditions in a large variety of systems. These dragon-kings reveal the existence of mechanisms of self-organization that are not apparent otherwise from the distribution of their smaller siblings. We present a generic phase diagram to explain the generation of dragon-kings and document their presence in six different examples (distribution of city sizes, distribution of acoustic emissions associated with material failure, distribution of velocity increments in hydrodynamic turbulence, distribution of financial drawdowns, distribution of the energies of epileptic seizures in humans and in model animals, distribution of the earthquake energies). We emphasize the importance of understanding dragon-kings as being often associated with a neighborhood of what can be called equivalently a phase transition, a bifurcation, a catastrophe (in the sense of René Thom), or a tipping point. The presence of a phase transition is crucial to learn how to diagnose in advance the symptoms associated with a coming dragon-king. Several examples of predictions using the derived log-periodic power law method are discussed, including material failure predictions and the forecasts of the end of financial bubbles.

## 1. Introduction

Systems with a large number of mutually interacting parts, often open to their environment, self-organize their internal structure and their dynamics with novel and sometimes surprising macroscopic ("emergent") properties. The complex system approach, which involves "seeing" inter-connections and relationships, i.e., the whole picture as well as the component parts, is nowadays pervasive in modern control of engineering devices and business management. It also plays an increasing role in most of the scientific disciplines, including biology (biological networks, ecology, evolution, origin of life, immunology, neurobiology, molecular biology, etc), geology (plate-tectonics, earthquakes and volcanoes, erosion and landscapes, climate and weather, environment, etc.), economy and social sciences (including cognition, distributed learning, interacting agents, etc.). There is a growing recognition that progress in most of these disciplines, in many of the pressing issues for our future welfare as well as for the management of our everyday life, will need such a systemic complex system and multidisciplinary approach. This view tends to replace the previous "analytical" approach, consisting of decomposing a system in components, such that the detailed understanding of each component was believed to bring understanding in the functioning of the whole.

One of the most remarkable emergent properties of natural and social sciences is that they are punctuated by rare large events, which often dominate their organization and lead to huge losses. This statement is usually quantified by heavy-tailed distributions of event sizes. Here, we present evidence that there is "life" beyond power laws: we introduce the concept of dragon-kings to refer to the existence of transient organization into extreme events that are statistically and mechanistically different from the rest of their smaller siblings. This realization opens the way for a systematic theory of predictability of catastrophes, which is outlined here and illustrated.

Section 2 reviews the evidence for power law distributions in many natural and social systems. Section 3 presents the limits of this power law description and documents the presence of dragon-kings in six different systems. Section 4 develops the concept that dragon-kings exhibit a degree of predictability, because they are associated with mechanisms expressed differently than for the other events. Often, dragon-kings are associated with the occurrence of a phase transition, bifurcation, catastrophe, tipping point, whose emergent organization produces useful precursors. A variety of concrete examples are described, especially on the application of the diagnostic of financial bubbles and the prediction of their demise. Section 5 concludes.

*\*also at the Department of Physics and at the Department of Earth Sciences, ETH Zurich*
  *E-mail address*: dsornette@ethz.ch.

## 2. Power law distributions of large event sizes in natural and social systems

Probability distribution functions with a power law dependence as a function of event or object sizes seem to be ubiquitous statistical features of natural and social systems. In complex systems, the appearance of power law distributions is often thought to be the signature of self-organizing mechanisms at the origin of a hierarchy of scales (see Sornette [1,2] for a general overview of power law distributions).

A probability distribution function P(x) exhibiting a power law tail is such that

$$P(x) \propto \frac{C_\mu}{x^{1+\mu}}$$

for large x, possibly up to some large limiting cut-off. The exponent μ (also referred to as the "index") characterizes the nature of the tail: for μ<2, one speaks of a "heavy tail" for which the variance is theoretically not defined.

The following two sub-sections provide a series of figures illustrating the occurrence of power law tails in both natural and social systems.

### 2.1. Power law distributions in natural systems

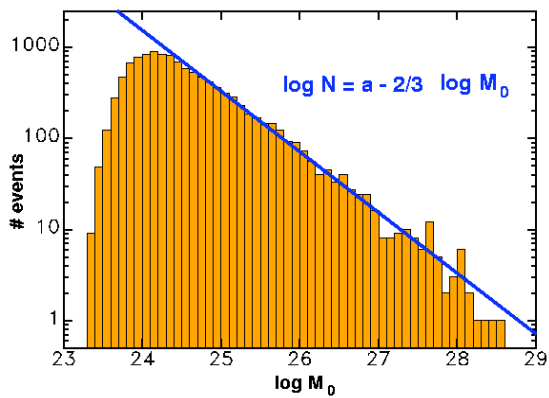

Fig.1 Distribution of earthquake seismic moments $M_0$ (a proxy roughly proportional to energy) in a large seismic-tectonic region such as California, USA. The straight line in log-log scale qualifies a power law distribution with exponent μ ≈2/3.

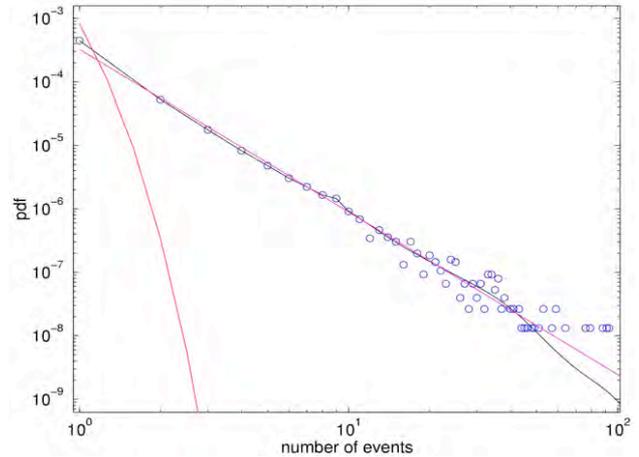

Fig.2 Distribution of earthquake seismic rates (number of events per day and per grid box) in Southern California, USA. The grid boxes are 5km x 5km in size. The straight line in log-log scale qualifies a power law distribution with exponent μ ≈2.5. The continuous red line is the best fit with a Poisson law, which is found completely off the data. Reproduced from Saichev and Sornette [3,4].

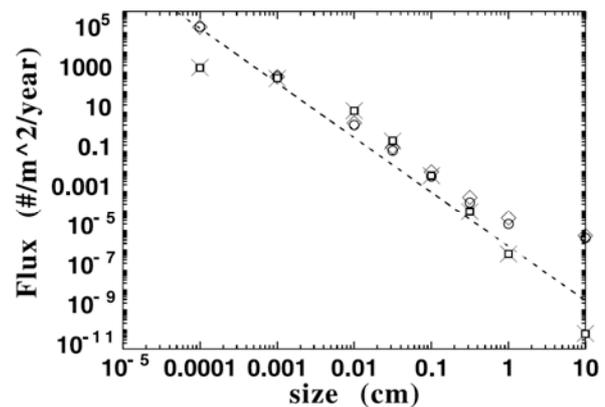

Fig.3 Distribution of meteorites and debris size orbiting around the Earth at different altitudes of 950km and 1500km above ground. The debris are man-made and are the remnants of rockets and satellites launched since the Soviet opened the space age with the launch of Sputnik I. The distributions are given in terms of the number of objects of a given size in centimeter crossing one square meter per year. The straight line in log-log scale qualifies a power law distribution with exponent μ ≈2.75. Reproduced from Sornette [2].

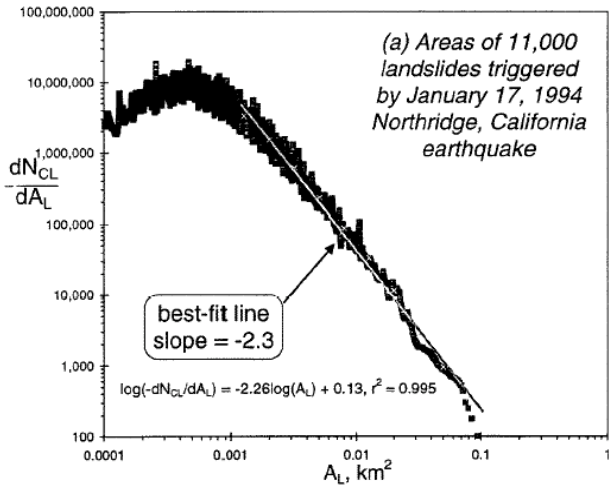

Fig.4 Distribution of areas of landslides triggered by the Jan. 17, 1994 Northridge earthquake, California, USA. The straight line in log-log scale qualifies a power law distribution with exponent µ ≈2/3. Reproduced from Turcotte [5].

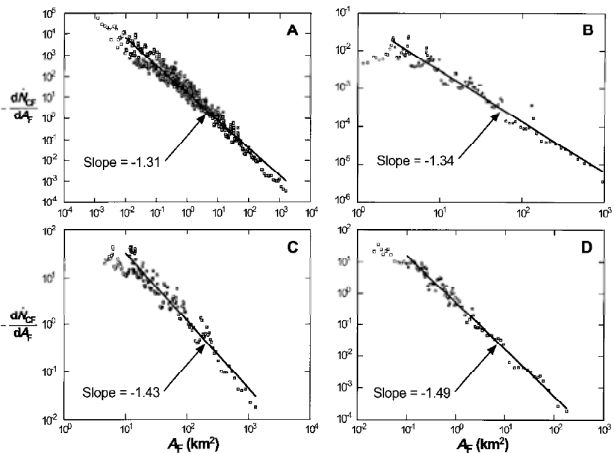

Fig.5 Distribution areas of forest fires and wildfires in the United States and Australia. (A) 4284 fires on U.S. Fish and Wildlife Service Lands (1986-1995); (B) 120 fires in the western United States (1150-1960); (C) 164 fires in Alaskan boreal forests (1990-1991); (D) 298 fires in the ACT (1926-1991). The straight line in log-log scale qualifies a power law distribution with exponent µ ≈0.3-0.5. Reproduced from Malamud et al. [6].

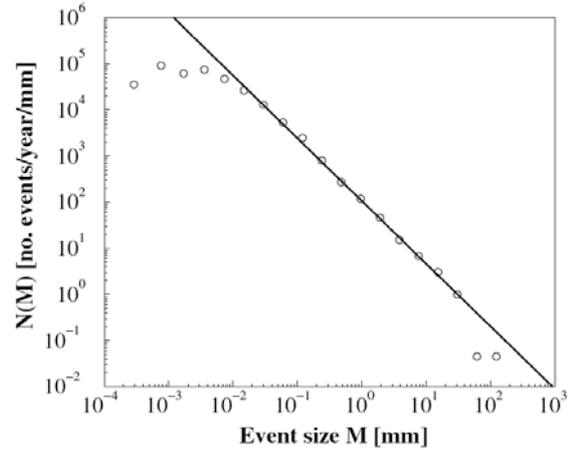

Fig.6 The number density N(M) of rain events versus the event size M (open circles) on a double logarithmic scale. Events are collected in bins of exponentially increasing widths. The horizontal position of a data point corresponds to the geometric mean of the beginning and the end of a bin. The vertical position is the number of events in that bin divided by the bin size. To facilitate comparison, the number of events are rescaled to annual values by dividing by the fraction of a whole year during which the data were collected. The straight line in log-log scale qualifies a power law distribution with exponent µ ≈0.4. Reproduced from Peters and Christensen [7].

## 2.2. Power law distributions in social systems

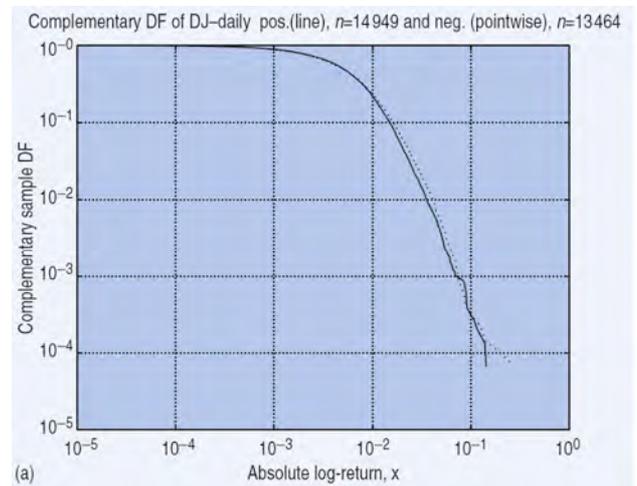

Fig.7 Survival distribution of positive (continuous line) and negative daily returns (dotted line) of the Dow Jones Industrial Average index over the time interval from May 27, 1896 to May 31, 2000, which represents a sample size of n=28 415 data points. The straight part in the tail in this log-log scale qualifies a power law distribution with exponent µ≈3. Reproduced from Malevergne et al. [8].

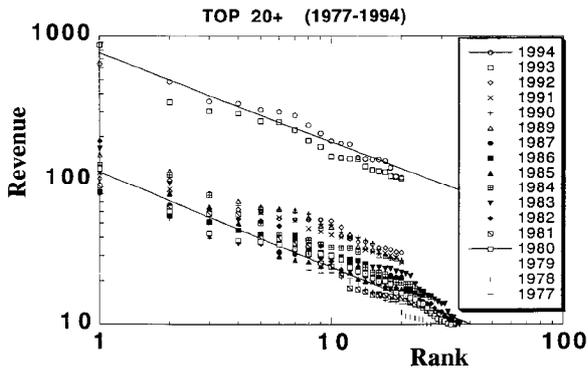

Fig.8 Rank-ordering plot of the revenue as a function of rank for the 50 top movies produced in Hollywood studios in terms of their revenues for each year from 1977 to 1994. The rank-ordering plot gives the same information as the empirical survival distribution, which derives from it by inverting the axes. The straight part in the tail in this log-log scale qualifies a power law distribution with exponent $\mu \approx 1.5$. Reproduced from Sornette and Zajdenweber [9].

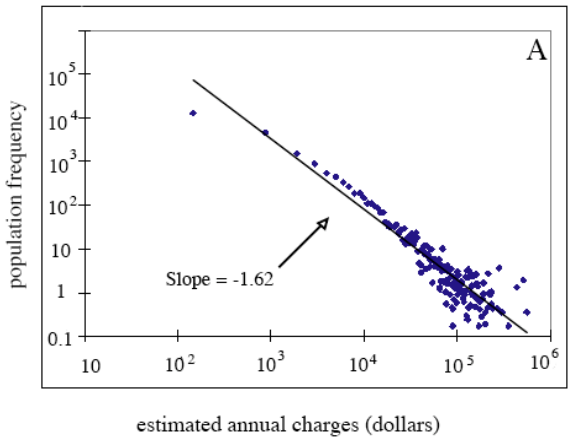

Fig.9 Population frequency of estimated health charges: Annual charge estimates from a 1998 U.S. national sample of non-institutionalized individuals. Data collected as part of the 1998 Medical Expenditure Panel Survey (MEPS) by the Agency for Healthcare Research and Quality. Annual health charges (excluding drug costs) ranged from none by 21% of participants, to a maximum of over $558,000. The straight part in the tail in this log-log scale qualifies a power law distribution with exponent $\mu \approx 0.6$. Reproduced from Rupper [10].

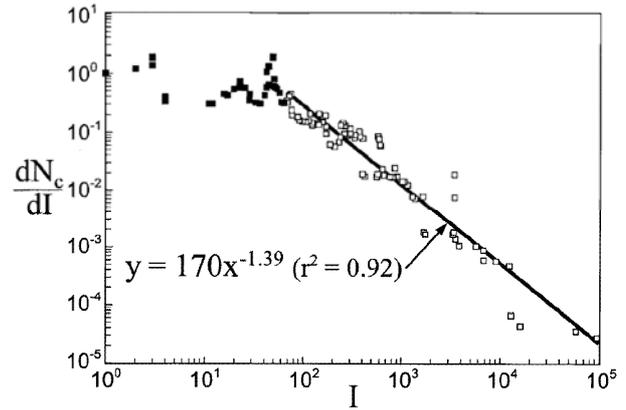

Fig.10 Frequency–intensity distribution of wars based on the Levy [11] tabulation of war intensities. The straight part in the tail in this log-log scale qualifies a power law distribution with exponent $\mu \approx 0.4$. Reproduced from Turcotte [5].

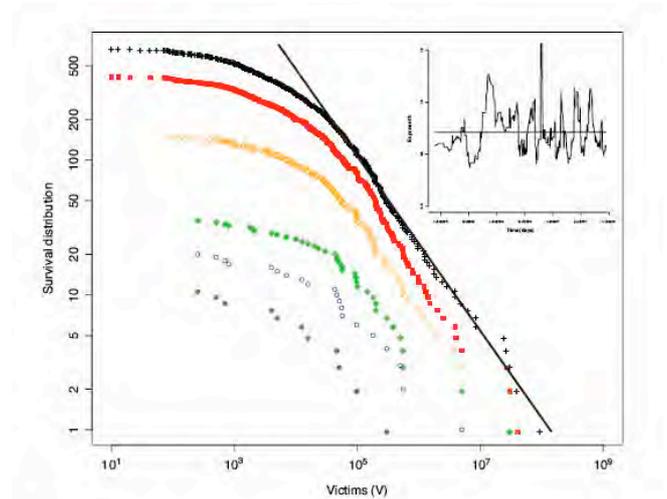

Fig.11 Non-normalized survival distribution (double logarithmic scale) of identity theft (ID) losses, constructed using the data provided in datalossdb.org/ (06.01.2009), respectively for events which have occurred before 2003 (filled purple circles), 2004 (empty blue circles), 2005 (green filled diamonds), 2006 (orange empty diamonds), 2007 (red squares) and the complete data set until December 20, 2007 (black crosses). Year after year, the tails of the survival distributions have approximately the same slope. The straight black line is the fit with a power law $\mu \approx 0.7$ for number of victims larger that the lower threshold $u=7 \cdot 10^4$. The inset shows the dependence of the index $\mu$ as a function of u obtained directly from the maximum likelihood estimation of the exponent of the power. Reproduced from Maillart and Sornette [12].

## 2.3. The standard view: tail events and black swans

Power law distributions incarnate the notion that extreme events are not exceptional events. Instead, extreme events should be considered to be rather frequent and to result from the same organization principle(s) as those generating other events: because they belong to the same statistical distribution, this suggests common generating mechanism(s). In this view, a great earthquake is just an earthquake that started small ... and did not stop; it is inherently unpredictable due to its sharing of all the properties and characteristics of smaller events (except for its size), so that no genuinely informative precursor can be identified [13]. If events of large impacts are part of a population described by a power law distribution, the common wisdom is that there is no way to predict them because nothing distinguish them from their small siblings: their great sizes and impacts come out as surprises, beyond the realm of normal expectations. This is the view expounded for instance by Bak and co-workers in their formulation of the concept of self-organized criticality [14,15]. This is also the concept espoused by the "Black Swan Theory" [16], which views high-impact rare events as unpredictable.

## 3. Beyond power laws: dragon-king outliers

Are power laws really the whole story? The following examples suggest that, in a significant number of complex systems, extreme events are even "wilder" than predicted by the extrapolation of the power law distributions in their tail. Below, we document evidence for what can be termed genuine "outliers" or even better "kings" [17] or "dragons."

According to the definition of the Engineering Statistical Handbook [18], "An outlier is an observation that lies an abnormal distance from other values in a random sample from a population." It is therefore an anomaly, an event to be removed in order to obtain reliable statistical estimations. The term "outlier" emphasizes the spurious nature of these anomalous events, suggesting to discard them as errors, or as misleading monsters.

In contrast, the term "king" has been introduced by Laherrère and Sornette [17] to emphasize the importance of those events, which are beyond the extrapolation of the fat tail distribution of the rest of the population. This is in analogy with the sometimes special position of the fortune of kings, which appear to exist beyond the Zipf law distribution of wealth of their subjects, as exemplified by King Buhimol Adulyadej (Thailand), Sheikh Khalifa bin Zayed al-Nahayan (United Arab Emirates), Sultan Hassanal Bolkiah (Brunei), Sheikh Mohammed Bin Rashid al-Maktoum (Dubai), Prince Hans Adam II (Liechtenstein), Sheikh Hamad bin Khalifa al-Thani (Qatar), King Mohammed VI (Morocco), Prince Albert II (Monaco) and so on [19]. I also like to refer to these exceptional events as "dragons" to stress that we deal with a completely different kind of animal, beyond the normal, which extraordinary characteristics, and whose presence, if confirmed, has profound significance.

The following sub-sections present empirical evidence of the presence and importance of dragon-kings in six different systems. A very important message is that there is no unique methodology to diagnose dragon-kings. One needs a battery of tools. Dragon-kings can be observed sometimes directly, in the form of obvious breaks or bumps in the tail of size distributions as in the example of sections 3.1 and 3.2. Or they need the construction of novel observables, which are more relevant to the dynamics of the system, as in the example of section 3.3. Or it is the comparison of distributions obtained at different resolution scales that allows one to diagnose the existence of a population of dragon-kings, as shown in the example of section 3.4. Section 3.5 demonstrates yet another mechanism for the generation of dragon-kings, found in the strong coupling regime of coupled heterogeneous oscillators of relaxation. A general phase diagram is presented which is tested on the statistics of epileptic seizures. Section 3.6 discusses the evidence supporting the predictions of the phase diagram for earthquake statistics. In this context, the dragon-kings would correspond to so-called "characteristic earthquakes."

### 3.1. Paris as the dragon-king of the Zipf distribution of French city sizes

Since Zipf's famous book [20], it is well documented that the distribution of city sizes (measured by the number of inhabitants) is, in many countries, a power law with an exponent $\mu$ close to 1. This ubiquitous regularity is understood as due to the law of proportional growth, also called Gibrat's law (see Saichev et al. [21] for a recent in-depth review of theories of Zipf's law and of its deviations). France is not an exception as it exhibits a nice power law distribution of city sizes... except that its capital, Paris, is completely out of range, and constitutes a genuine dragon-king with a size several times larger than expected from the distribution of the rest of the population of cities [17]. This phenomenon is represented in Fig.12, showing a rank-ordering plot of the sizes S of French cities (raised to the exponent c=0.18) as a function of the logarithm of the city rank, ordered by descending sizes. This representation qualifies a stretched exponential distribution, which takes the form $\sim \exp[-(S/S_0)^c]$ [4,16]. Malevergne et al. [8] have shown that power law distributions are embedded as special cases in the large family of stretched exponential distributions in the following sense: a stretched exponential distribution degenerates into a power law distribution with finite exponent $\mu$ in the limit when the exponent c and the characteristic scale $S_0$ becomes much smaller than 1, while preserving the finite limit $c (u/S_0)^c \Rightarrow \mu$, where u is the lower threshold above with the power law holds. In this sense, because the best empirical stretched exponential exponent c=0.18 is small, the straight line in Fig.12 qualifies a power law distribution, with an exponent µ which turns out to be close to 1 (Zipf's law).

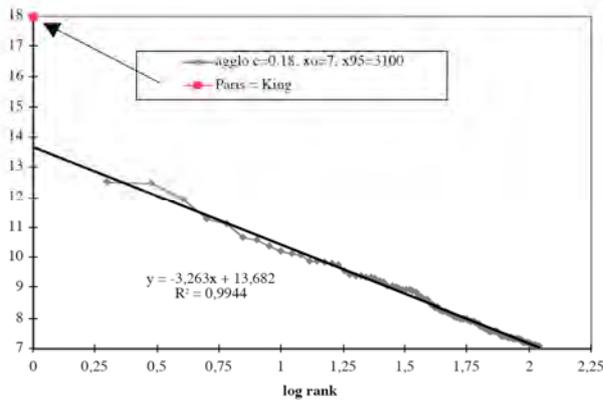

Fig.12 Rank-ordering plot of the population size of French cities as a function of their rank, where sizes are ordered in decreasing values. The rank-ordering plot gives the same information as the empirical survival distribution, which derives from it by inverting the axes. For the ordinate, the city size is raised to the power c=0.18 and the abscissa represents the logarithm of the rank. A straight line then qualified a stretched exponential distribution $\sim \exp[-(S/S_0)^c]$. See text for how this qualifies a power law. The arrow shows the data point for Paris. Reproduced from Laherrère and Sornette [17].

But the most interesting message of Fig.12 is the exceptional deviation of a single point, rank 1, which is Paris, the largest city of France. Clearly, this city does not abide to Zipf's power law.

Do we want to throw away or neglect this information as a spurious outlier? Should we ignore the role of Paris in the distribution of French city sizes? Actually, as is well known, Paris has played historically a crucial role in the development of France, and its dragon-king status observed here in the statistical distribution of French city sizes is a revealing sign of this rich and complex history. We will show in the following examples that the dragon-king status emerges in general from the existence of positive feedbacks, that amplify the role of certain events. In the case of Paris, the centralized organization for French governments over the past centuries has led to its ever-increasing pivotal role. London plays a similar dragon-king role with respect to the distribution of British city sizes. This evidence provides a clue that the existence of a dragon-king is associated with special mechanisms of amplifications.

**3.2. Global failure as the dragon-king in material failure and rupture processes**

There is now ample evidence that the distribution of damage events, for instance quantified by the acoustic emission radiated by micro-cracking in heterogeneous systems, is well-described by a Gutenberg-Richter like power law [22-25]. But consider now the energy released in the final global event rupturing the system in pieces, as shown in Fig.13. This release of energy is many times larger than the largest ever recorded event in the power law distribution before the occurrence of the run-away rupture. Material rupture exemplifies the co-existence of a power law distribution and a catastrophic dragon-king event lying beyond the power law.

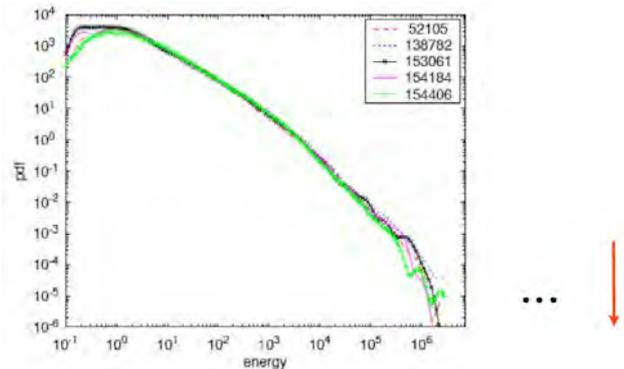

Fig.13 Distribution of acoustic emissions recorded in creep experiments of composite materials subjected to a constant stress. The different curves correspond to the distributions obtained at different epochs along the lifetime of the sample, confirming the stability and robustness of the distributions. The approximately linear behaviors observed in the tails in this log-log plot qualify power law distributions. The vertical red arrow illustrates the point made in the text that the final rupture associated with a run-away crack releases much more energy, as can be heard by the corresponding snapping sound. The plot is reproduced from Nechad et al. [26].

The main positive feedback mechanisms at the origin of the run-away dragon-king occurring in material failure have been gathered by Sammis and Sornette [27].

**3.3. Dragon-kings in the distribution of financial drawdowns (or run of losses)**

Fig.7 shows the survival distribution of positive (continuous line) and negative daily returns (dotted line) of the Dow Jones Industrial Average index over the time interval from May 27, 1896 to May 31, 2000 [7]. No dragon-king is apparent and it seems that the distribution of large losses and large gains are pure asymptotic power laws [28].

But this is missing the forest for the tree! Our claim is that financial returns defined at fixed time scales, say at the hourly, daily, weekly or monthly time scales, are revealing only a part of the variability of financial time series, while a major risk component is gravely missing.

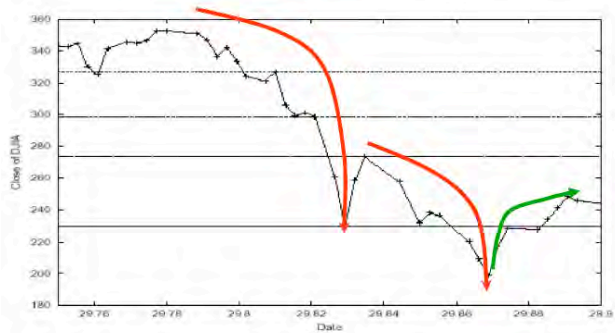

Fig.14 A short time series of the DJIA on which the red arrows define two drawdowns, while the green arrow defines a drawup. There are many possible definitions of drawdowns (and their symmetric drawups). Details are given in Johansen and Sornette [29-31].

Since we are interested in characterizing the statistics of extreme events, to illustrate our claim, consider the simplified textbook example of a crash that occurs over a period of three days, as a sequence of three successive drops of 10% each summing up to a total loss of 30%. Now, a 10% market drop, that occurs over one day, can be seen in the data of the Nasdaq composite index to happen on average once every four years. Since there are approximately 250 trading days in a year, a 10% market drop is thus a $10^{-3}$ probability event. What is the probability for observing three such drops in a row? The answer is $(10^{-3})^3=10^{-9}$. Such one-in-one-billion event has a recurrence time of roughly 4 million years! Thus, it should never be observed in our short available time series. However, many crashes of such sizes or larger have occurred in the last decades all over the world.

What is wrong with the reasoning leading to the exceedingly small $10^{-9}$ probability for such a crash? It is the assumption of independence between the three losses! In contrast, our claim is that financial crashes are transient bursts of dependence between successive large losses. As such, they are missed by the standard one-point statistics consisting in decomposing the runs of losses into elementary daily returns. With some exaggeration to emphasize my message, I would say that by cutting the mammoth in pieces, we only observe mice.

We thus propose to analyze drawdowns (and their symmetrical drawups), because they are better adapted to capture the risk perception of investors, and therefore better reflect the realized market risks. Indeed, we demonstrate below that the distributions of drawdowns diagnose efficiently financial crashes, which are seen as dragon-kings, i.e., special events associated with specific bubble regimes that precede them.

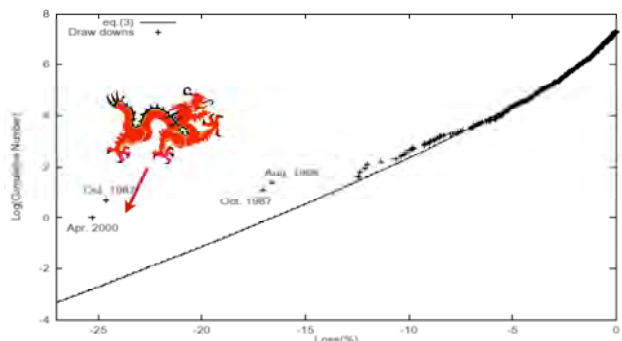

Fig.15 Distribution of drawdowns D for the Nasdaq Composite index, showing several "outliers" in the tail, that qualify as dragon-kings. It turns out that these anomalous events are precisely associated with documented crashes, such at the crash of the ITC bubble in April 2000 or the Oct. 1987 World stock market crash. The continuous line is the fit to the stretched exponential distribution $\sim \exp[-(D/D_0)^c]$ with an exponent $c \approx 0.8$. Reproduced from Johansen and Sornette [30].

Fig.15 show the distribution of drawdowns obtained from the Nasdaq composite index over a 20-year period, which includes several great crashes shown by the arrow. As analyzed carefully by Johansen and Sornette [29-32], about 99% of the drawdowns can be represented nicely by a common distribution, which turns out to have a tail slightly fatter than an exponential. And the remaining few events have been found to be statistically different: the hypothesis that they belong to the same distribution as 99% of the population of the other drawdowns is rejected at the 99.9% confidence level [32,33]. Fig. 16 presents further evidence for 30 individual US companies of the existence of dragon-kings in the distribution of drawdowns (run of losses) and drawups (runs of gains).

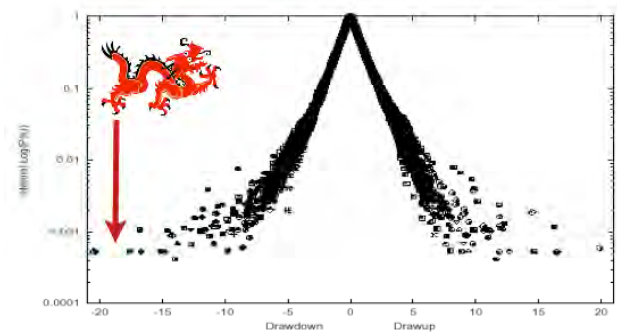

Fig.16 Same as Fig 15 for 30 major US companies. This figure shows in addition the distribution of drawups, i.e., runs of gains, depicted on the right side of the peak. The drawdown/drawup distribution of each company has been scaled by its corresponding scale factor $D_0$ defined in the stretched exponential fit mentioned in Fig.15, so as to be able to superimpose the 30 distribution functions. The collapse of the 30 curves is good for the bulk but fails in the tails, where dragon-kings can be observed.

The evidence of dragon-kings reported by Johansen and Sornette [29-33] encompasses exchange markets (US dollar against the Deutsch Mark and against the Yen), the major world stock markets, the U.S. and Japanese bond markets and commodity markets. The results have been found robust with respect to change in various measures of drawdowns, in particular which allow for a certain degree of fuzziness in the definition of cumulative losses [31,32].

Johansen and Sornette [31] have found that two-thirds of the dragon-kings identified in the distribution of drawdowns are actually stock market crashes, which were preceded by large bubbles. This has led them to build a theory [33-35] in which crashes are seen as the possible end of a bubble regime, associated with various positive feedback mechanisms that lead to faster-than-exponential unsustainable growth regime. The mechanisms responsible for positive feedbacks include portfolio insurance trading, option hedging, momentum investment and imitation-based herding.

### 3.4. Dragon-king events in the distribution of turbulent velocity fluctuations.

Until now, we have emphasized that dragon-kings can be identified as extreme outliers in the tail of the distribution of event sizes, and correspond to some kind of break or bump. Actually, dragon-kings do not always lead to this diagnostic and other measures are necessary to identify their presence.

This point is well illustrated in shell models of turbulence, that are believed to capture the essential ingredient of these flows, while being amenable to quantitative analysis. Such shell models replace the three-dimensional spatial domain by a series of uniform onion-like spherical layers with radii increasing as a geometrical series $1, 2, 4, 8, ..., 2^n$ and communicating mostly with nearest neighbors. The quantity of interest is the distribution of velocity variations between two instants at the same position or between two points simultaneously. L'vov et al. [36] have shown that they could collapse the distribution function of velocity fluctuations for different scales only for the small velocity fluctuations, while no scaling held for large velocity fluctuations, as shown in Fig. 17.

Fig.17 suggests that the distributions of velocity fluctuations are composed of two regions, a part corresponding to so-called normal scaling and a domain of extreme events. The extreme events can actually be visualized directly as they correspond to intense peaks propagating coherently (like solitons) over several shell layers with a characteristic bell-like shape, approximately independent of their amplitude and duration (up to a rescaling of their size and duration). The two coexisting populations correspond to "characteristic" velocity pulses decorating incoherent scale-invariant velocity fluctuations.

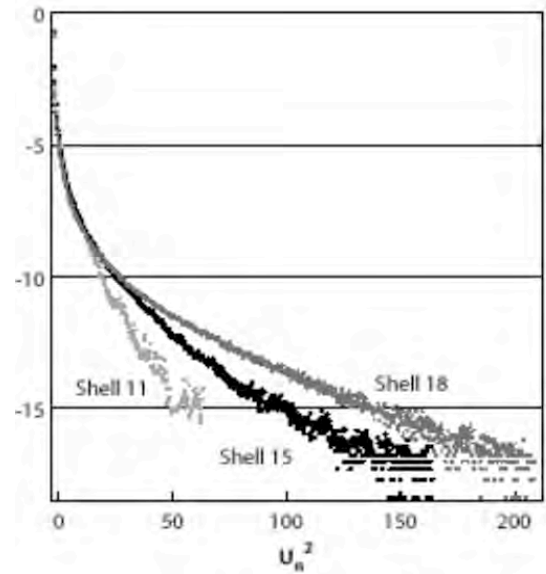

Fig.17 Distribution of the square of the velocity in three different shells, corresponding to three different spatial scales. The hypothesis, that there are no outliers, is tested here by collapsing the distributions for the three shown layers. While this is possible for small velocities, the tails of the distributions for large events are very different in the three shells, indicating that extreme fluctuations belong to a different class of their own. Reproduced from L'vov et al. [36].

This example emphasizes that the diagnostic of dragon-kings requires different methods adapted to the specific problem. Here, the identification of the dragon-kings relies on the comparison between distributions of event sizes obtained at different resolution scales.

### 3.5. Dragon-kings in distributions of epileptic seizures associated with the strong coupling synchronized regime

Resulting from a neurological disorder that affects 60 million humans worldwide, epileptic seizures are typically associated with marked paroxysmal increases in the amplitude or rhythmicity of neuronal oscillations which, in a large number of subjects, begin in a discrete region of the brain, but may eventually spread to engulf the entire brain.

Osorio et al. [37,38] have recently reported that the statistics of epileptic seizures in human subjects and in animal models closely resembles that observed for earthquakes: power laws govern the distributions of seizure energies and of recurrence times between events; moreover, the rates of seizures prior and posterior to a seizure follow respectively the so-called inverse and direct Omori power laws.

The close correspondence in four different statistics observed between seizures and earthquakes can be traced conceptually [37,38] to the fact that both types of events are generated by systems composed of interacting (coupled) relaxation threshold oscillators.

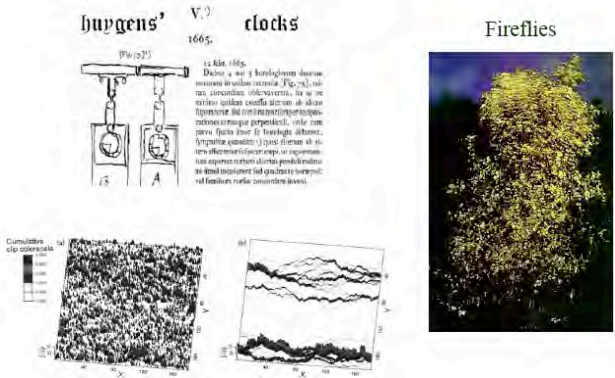

Fig.18 Three examples of coupled oscillators which tend to exhibit the phenomenon of synchronization. The upper left panel represents the famous historical example of two clocks coupled via the vibrations emitted from their pendula and transmitted through the wall on which they were fixed, and which led Huygens to discover the phenomenon of synchronization of oscillators (clocks). The figure on the right shows a tree covered with male fireflies in Thailand, which synchronize their light signal under the mutual coupling via perception of neighbors flashing (Reproduced from Buck and Buck [39]). The bottom left panel represents a model of seismic faults that are coupled via elastic stresses. This model was studied by Sornette et al. [40] and led to propose the general phase diagram shown in Fig.20.

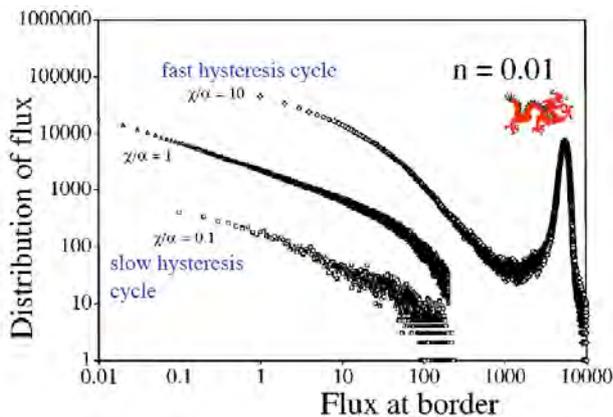

Fig.19 Distribution P(J) of flux amplitudes at the right border of system described as a Landau-Ginzburg model sandpile. The model uses a continuous framework with a noisy nonlinear diffusion equation controlling the space-time evolution of the control parameter (slope of the sandpile), which is coupled to the order-parameter (state of rolling of sand grains) described by the normal form of a sub-critical bifurcation. $\chi/\alpha$ is the ratio of the two characteristic time scales of the problem, the time scale associated with sand diffusion over the time scale of the transition from static to rolling described by the order parameter. The amplitude n of the driving noise corresponds to the small noise regime. Reproduced from Gil and Sornette [41].

A generic phase diagram [40] shown in Fig.19 depicts the main different regimes exhibited by systems made of heterogeneous coupled threshold oscillators, such as sandpile models [41] (see Fig.19), integrate-and-fire oscillators [42], financial market models [43], Burridge-Knopoff block-spring models [44] and earthquake-fault models [45]: a power law regime (probably self-organized critical) (Fig.20, right lower half) is co-extensive with one of synchronization with characteristic size events (Fig.20, upper left half). Synchronization herein refers to a coherent dynamics of coupled oscillators, and does not necessarily require unison or simultaneous beating.

The appearance of characteristic dragon-king events in the distribution of forest fires is also characteristic of forest-fire cellular automata models in the limit where the sparking frequency goes to zero (see the review and figures in section 4 in Turcotte [5]). Bumps in the distribution of large returns occur when a measure of coupling between investors increases above some critical threshold [43].

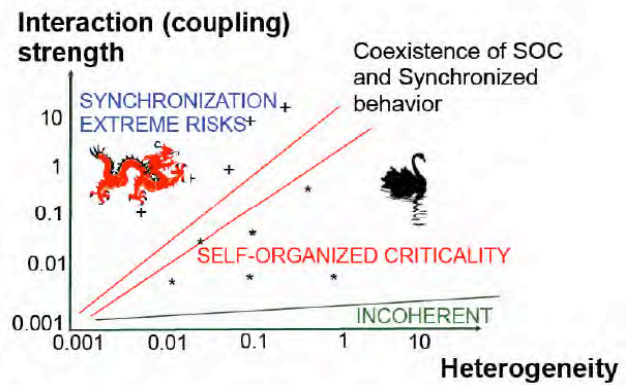

Fig.20 Qualitative phase diagram illustrating the effect of changes in coupling strength (y-axis) and heterogeneity (x-axis) on the behavior of systems (such as the brain and the Earth crust), composed of interacting threshold oscillators (only changes in coupling were investigated in animal models for epileptic seizures [37,38]). Marked increases in excitatory coupling (high 3-MPA dose) drive the system towards the synchronized regime tagged by the dragon-king. Slight increases in coupling (low 3-MPA dose) drive the system towards the power law regime indicative of self-organized criticality, with the representative black swan icon. Reproduced from Sornette [45] and Osorio et al. [38].

As shown in this diagram, when coupling is weak (and/or heterogeneity is strong), the power law regime prevails (self-organized criticality (SOC) regime); as the coupling strength increases (and/or heterogeneity decreases), the systems moves towards the synchronization regime and events occurs periodically. The black swan cartoon stresses the fact that, in the SOC regime, the extreme events are no different from their smaller siblings, making the former unpredictable. This is the power law regime described in section 2.3. In contrast, the dragon-king icon stresses the fact that the extreme events occurring in the synchronized regime are different for the vast majority of the population, i.e., they constitute anomalies, in the sense illustrated by the different examples shown in the present section 3.

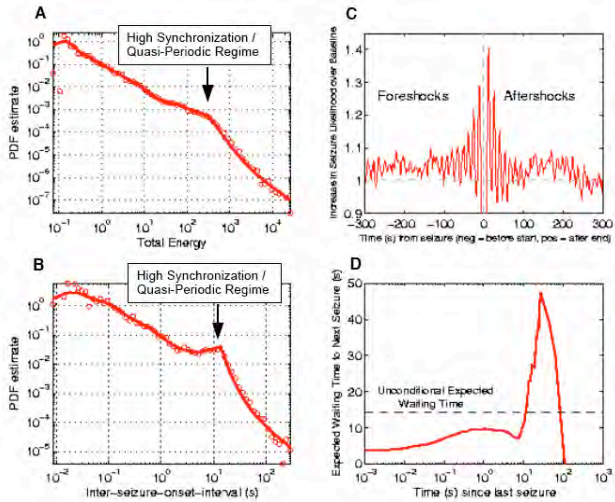

Fig.21 Top-left panel: distribution of seizure energies; top-right panel: superimposed epoch analysis of seizures to test for the existence in seizures of "aftershocks" (Omori-like) and "foreshocks" (inverse Omori-like law); bottom-left panel: distribution of inter-seizure waiting times; bottom-right panel: conditional average waiting time till the next seizures as a function of the time spent since the last event. These statistics are obtained for rats treated with a convulsant substance (3-MPA). Reproduced from Osorio et al. [38].

The generic phase diagram shown in Fig.20 leads to the prediction that, if one can control the degree of heterogeneity and/or the level of coupling between the coupled threshold oscillators, one should be able to observe a change of regime from SOC to synchronized or vice-versa. While one cannot easily manipulate either parameter at the scale of the Earth crust, one can control the level of coupling for seizures, taking as a proxy for it the amount of epileptogenic substances injected in animals. This was done by Osorio et al. [38] who tested, in an animal seizure model, the earthquake-driven hypothesis that power law statistics co-exist with characteristic scales, as coupling between constitutive elements increases towards the synchronization regime.

Specifically, the seizure energy distribution of 9 rats treated with relatively low 3-MPA convultant doses, so as to only cause a moderate increase in excitatory neuronal coupling (relative to untreated rats), a state that corresponds to the weak coupling power regime (lower half, Fig.20) of the generic phase diagram, follows a power law distribution. In contrast, the seizure energy distribution of 19 rats that were treated with maximally tolerable (for viability) steady-state brain concentrations of 3-MPA, a state that corresponds to the strong coupling regime (Fig.20, upper half), had power law behavior (2 decades on the x- and 3 on the y-axis), coextensive with characteristic scales (Fig. 21, left upper panel). High 3-MPA concentrations in brain induced very frequent, prolonged seizures, that violated the linear regime in log-log scale, forming a "shoulder" (arrow, Fig.21 left upper panel), indicative of a characteristic seizure size.

Quasi-periodic behavior is also clearly seen in the seizure "foreshock/aftershock" plots (Fig.21 right upper panel) in the shape of regularly spaced oscillations "decorating" the inverse and direct Omori laws. The distribution of inter-seizure intervals also exhibits a clear characteristic time scale (arrow, Fig.21 lower left panel). Correspondingly, the average conditional waiting time (Fig.21 right lower panel) is also highly suggestive of quasi-periodic behavior superimposed on some large waiting time occurrences.

These results of manipulating the strength of excitatory inter-neuronal coupling with 3-MPA furnishes evidence in support of the concepts illustrated in the generic phase diagram (Fig.20). Modest increases in coupling strength manifest as scale-free events, which are likely the expression of SOC (or perhaps better expressed as "critical asynchronization" [46]) while marked increases generate events with characteristic scales (i.e., periodic), advocating yet another prediction: Seizures with characteristic scales should also be observable in humans, as their epileptogenic brain explores the strong excitatory coupling state. This prediction has been confirmed by Osorio et al. [38] (not shown here).

This buttresses the argument that, in animals and humans, scale-free is not the only behavior of systems populated by relaxation threshold oscillators (neurons in this case). In particular, increases in interneuronal excitatory coupling generate characteristic scale seizures regimes that co-exist in space-time with scale-free ones. More generally, the theory underlying the correspondence between seizures and earthquakes implies that wide spectra of different dynamic regimes are possible for systems such as the brain's cortex and the earth's crust. These regimes could correspond to critical asynchrony/self-organized criticality [46], clustering, quasi-periodicity, and/or synchronization, depending on the convulsant concentration, its rate of change, and other physico-chemical changes in the neuropil (engendered by SZ) that may be likened to changes in soil structure/composition/water content associated with earthquakes.

### 3.6. Gutenberg-Richter law and characteristic earthquakes.

As mentioned above, earthquakes can be thought of as relaxation events of coupled heterogeneous faults, each fault acting as a threshold oscillator of relaxation under the influence of an overall slow tectonic loading. Given the analogy between earthquakes and seizures, the results, predicted by the phase diagram presented in Fig.20 and verified on strongly "coupled" rats' brains, suggest extrapolating from seizures to earthquakes. Specifically, Osorio et al. [38] have suggested that the controversial characteristic earthquake hypothesis [47-52] could correspond to a model of seismicity that should be observed only when coupling between faults is strong and heterogeneity is weak.

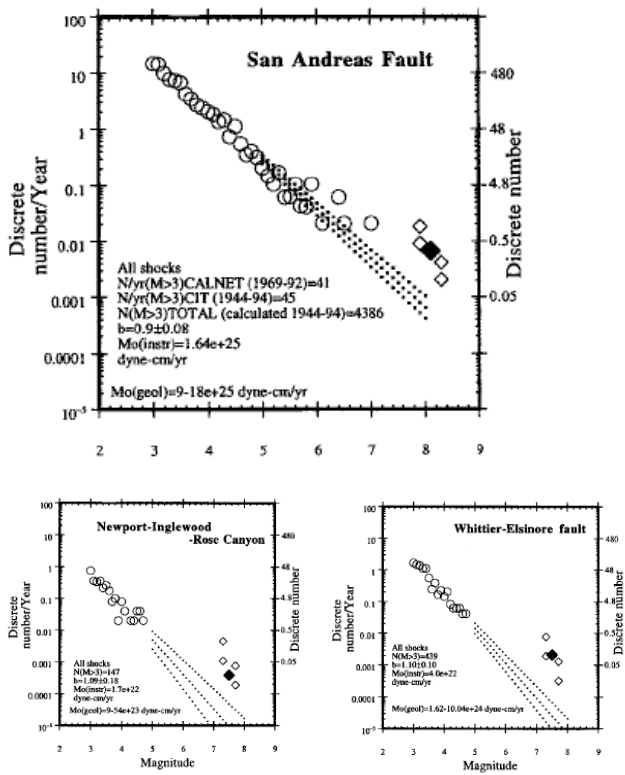

Fig.22 Top panel: distribution of earthquake magnitudes in a thin strip around the San Andreas fault in California. Bottom left and right panels: same as the top panel for the Newport-Inglewood-Rose Canyon fault and the Whittier-Elsinore fault, respectively. The characteristic earthquakes (dragon-kings) would be associated with the clusters, which are visibly above the extrapolation of the Gutenberg-Richter distribution calibrated on the smaller events. Reproduced from Wesnousky [49].

Testing for the dragon-king characteristic earthquake is difficult in the seismo-tectonic context, because of the difficulties with defining unambiguously the spatial domain of influence of a given fault over which the relevant statistics should be defined. The researchers who have delineated a spatial domain surrounding a clearly mapped large fault claim to find a Gutenberg-Richter distribution up to a large magnitude region characterized by a bump or anomalous rate of large earthquakes, as illustrated in Fig.22. These large "characteristic" earthquakes have rupture lengths comparable with the fault length [48,49]. If proven valid, this concept of a characteristic earthquake provides another example in which a dragon-king coexists with a power law distribution of smaller events. Others have countered that this bump disappears when removing the somewhat artificial partition of the data [50,51], so that the characteristic earthquake concept might be a statistical artifact. In this view, a particular fault may appear to have characteristic earthquakes, but the stress-shedding region, as a whole, behaves according to a pure scale-free power law distribution. Aki [52] presents a balanced view of this thorny issue of the existence or absence of characteristic earthquakes. The present discussion contributes by recognizing the relevance of material properties in shaping diverse possible seismicity regimes and calls for the re-examination of the characteristic earthquake hypothesis from this novel perspective.

Actually, several theoretical models have been offered to support the idea that, in some seismic regimes, a power law distribution earthquake energies coexists with a characteristic earthquake regime (the dragon-king effect). Gil and Sornette [41] reported that this occurs when the characteristic rate for local stress relaxations is fast compared with the diffusion of stress within the system. The interplay between dynamical effects and heterogeneity has also been shown to change the Gutenberg-Richter behavior to a distribution of small events combined with characteristic system size events [53-56]. Huang et al carried [57] out simulations on a square array of blocks using static-dynamic friction and a cellular-automata approach. Their frequency–area density distribution statistics for model slip events also exhibit the coexistence of a power law and a bump associated with catastrophic slip events involving the entire system.

On the empirical side, progress should be made in testing the characteristic earthquake hypothesis by using the prediction of the models to identify independently of seismicity those seismic regions in which the dragon-king effect is expected. This remains to be done [Ben-Zion, private communication, 2007].

## 4. Consequences of the dragon-king phenomenon for the predictability of catastrophic events

The fact, that dragon-kings belong to a statistical population, which is different from the bulk of the distribution of smaller events, requires some additional amplification mechanisms involving amplifying critical cascades active only at special times. In 2002, I have presented a preliminary review [58] of the methods based on these insights to predict material rupture, turbulence bursts, abrupt changes in weather regimes, financial crashes and human birth.

The key idea is that catastrophic events involve interactions between structures at many different scales that lead to the emergence of transitions between collective regimes of organization. One of the most important concepts developed in the theory of complex systems and in statistical physics is that big disruptions do not need large perturbations to occur. Most complex systems of interest exhibit qualitative changes of regimes in their characteristics and dynamics upon the smooth variations of some "control" parameters or as a function of the network topology and/or metric. These qualitative changes are known under a variety of names, such as ruptures, phase transitions, bifurcations, catastrophes, tipping points. These are often at the source of the dragon-kings describe above. These bifurcations take engineers, practitioners and students by surprise, because of the ubiquitous tendency to extrapolate new behavior from past ones. Such inferences are fundamentally mistaken at phase transitions, since the new collective organization is in general completely different from the previous one. It is also wrongly considered as unrelated.

Neglecting the fundamental relationships between dragon-kings and the pre-existing regimes would be a conceptual error with serious practical consequences. Methods that recognize the role of phase transitions allow us to unify different regimes under a synthetic framework, sometimes with encouraging potential for prediction of crises [58]. I briefly present two examples, in the field of material science and in financial economics.

### 4.1. Prediction of material failure

Fig 23 shows four plots of the acoustic emission recorded as the stress is increased linearly with time on pressure tanks embarked on rockets, which are made of multi-layer carbon composites. Using a theory of positive feedback of the present damage that impacts future damage [27,59,60], one can predict the final rupture to be a finite-time singularity with specific log-periodic power law (LPPL) precursors. The calibration of this model to data shows excellent results. This prediction system is now used routinely in the Aerospace industry in Europe to qualify the reliability of structures made of composite materials [61].

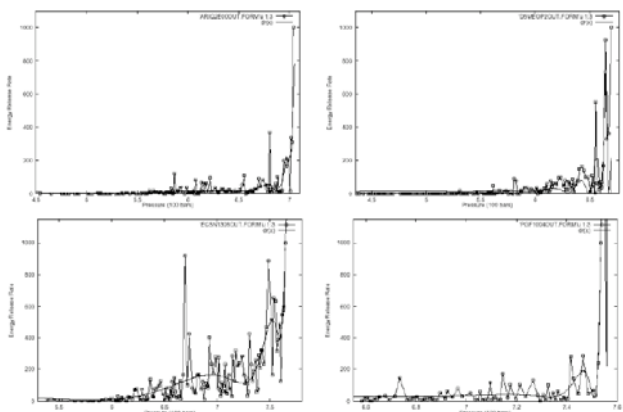

Fig.23 Illustration of the acoutic emission energy (vertical axis) recorded as the applied stress (horizontal axis) is increased on pressure tanks embarked on rockets, which are made of multi-layer carbon composites. Note the intermittent acceleration of the rate of acoustic emission energy that is well fitted by the LPPL model mentioned in the text. Reproduced from Johansen and Sornette [60].

### 4.2. Prediction of the end of financial bubbles

Stock market crashes are momentous financial events that are fascinating to academics and practitioners alike. According to the standard academic textbook worldview that markets are efficient, only the revelation of a dramatic piece of information can cause a crash, yet in reality even the most thorough post-mortem analyses are typically inconclusive as to what this piece of information might have been. For traders and investors, the fear of a crash is a perpetual source of stress, and the onset of the event itself always ruins the lives of some of them. Most approaches to explain crashes search for possible mechanisms or effects that operate at very short time scales (hours, days or weeks at most). Other researchers have suggested market crashes may have endogenous origins (see Kaizoji and Sornette [62] and references therein).

Associated with these questions is the problem of determining if there exist qualifying signatures in the statistical properties of time series of price returns that make crashes, and more generally large losses, different from the rest of the population? Section 3.3 has answered positively by showing that crashes are outliers or "dragon-kings" (in the sense of forming a different statistical population with extreme properties).

Financial markets constitute one among many other systems exhibiting a complex organization and dynamics with similar behavior. Over the last 15 years, we have developed an approach to challenge the standard economic view that stock markets are both efficient and unpredictable. The main concepts that are needed to understand stock markets are imitation, herding, self-organized cooperativity and positive feedbacks, leading to the development of endogenous instabilities. According to this theory, local effects such as interest raises, new tax laws, new regulations and so on, invoked as the cause of the burst of a given bubble leading to a crash, are only one of the triggering factors but not the fundamental cause of the bubble collapse. We propose that the true origin of a bubble and of its collapse lies in the unsustainable pace of stock market price growth based on self-reinforcing over-optimistic anticipation. As a speculative bubble develops, it becomes more and more unstable and very susceptible to any disturbance.

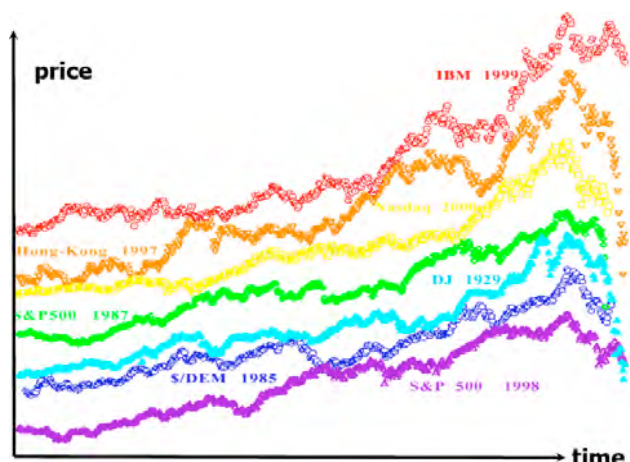

Fig.24 Seven well-known bubbles ending in a crash (not shown at the end of the time series) after an accelerated super-exponential increase fueled by positive feedback processes, in particular herding. Each time series has been rescaled vertically and translated to end at their corresponding crash time. The legends in color indicate the corresponding financial asset and the date is the year at which the bubble ended and the crash occurred.

In a given financial bubble, it is the expectation of future earnings rather than present economic reality that motivates the average investor. History provides many examples of bubbles driven by unrealistic expectations of future earnings followed by crashes. The same basic ingredients are found repeatedly. Markets go through a series of stages, beginning with a market or sector that is successful, with strong fundamentals. Credit expands, and money flows more easily. (Near the peak of Japan's bubble in 1990, Japan's banks were lending money for real estate purchases at more than the value of the property, expecting the value to rise quickly.) As more money is available, prices rise. More investors are drawn in, and expectations for quick profits rise. The bubble expands, and then bursts. In other words, fuelled by initially well-founded economic fundamentals, investors develop a self-fulfilling enthusiasm by an imitative process or crowd behavior that leads to the building of castles in the air, to paraphrase Malkiel [63]. Furthermore, the causes of the crashes on the US markets in 1929, 1987, 1998 and in 2000 belongs to the same category, the difference being mainly in which sector the bubble was created: in 1929, it was utilities; in 1987, the bubble was supported by a general deregulation of the market with many new private investors entering the market with very high expectations with respect to the profit they would make; in 1998, it was an enormous expectation with respect to the investment opportunities in Russia that collapsed; before 2000, it was extremely high expectations with respect to the Internet, telecommunications, etc., that fuelled the bubble. In 1929, 1987 and 2000, the concept of a "new economy" was each time promoted as the rational origin of the upsurge of the prices.

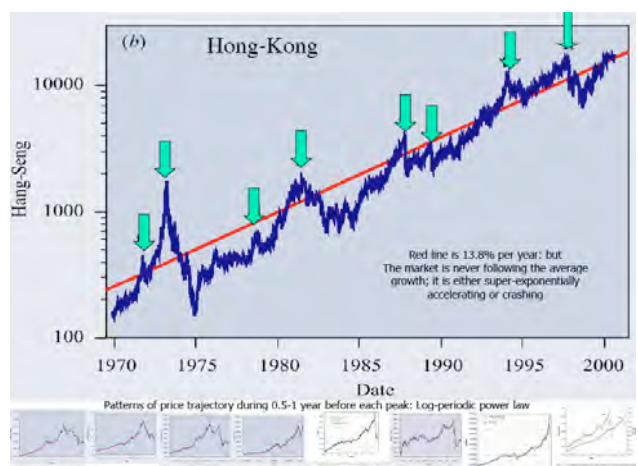

Fig.25 Thirty years of the value of the Hong-Kong Heng-Seng index (linear-log scale). The straight red line represents the long-term average exponential growth of the Heng-Seng index at an approximately constant growth rate of 13.8%. But what is more interesting is that this market is never following this average exponential trend: it is accelerating super-exponentially and then crashing in a succession of bubbles and corrections that can be observed to occur at several scales. The 8 arrows indicate the times when a market peak was followed by a drop of more than 15% in less than three weeks. The 8 small panels at the bottom show the market price over 1-2 years preceding each of these 8 peaks and the corresponding fits with the LPPL model discussed in the text. Reproduced from Sornette and Johansen [32].

Mathematically, large stock market crashes are the social analogues of so-called critical points studied in the statistical physics community in relation to magnetism, melting, and other phase transformation of solids, liquids, gas and other phases of matter [2]. This theory is based on the existence of a cooperative behavior of traders imitating each other which leads to progressively increasing build-up of market cooperativity, or effective interactions between investors, often translated into accelerating ascent of the market price over months and years before the crash. According to this theory, a crash occurs because the market has entered an unstable phase and any small disturbance or process may have triggered the instability.

Think of a ruler held up vertically on your finger: this very unstable position will lead eventually to its collapse, as a result of a small (or absence of adequate) motion of your hand or due to any tiny whiff of air. The collapse is fundamentally due to the unstable position; the instantaneous cause of the collapse is secondary. In the same vein, the growth of the sensitivity and the growing instability of the market close to such a critical point might explain why attempts to unravel the local origin of the crash have been so diverse. Essentially, anything would work once the system is ripe. In this view, a crash has fundamentally an endogenous or internal origin and exogenous or external shocks only serve as triggering factors.

As a consequence, the origin of crashes is much more subtle than often thought, as it is constructed progressively by the market as a whole, as a self-organizing process with universal properties, as illustrated in Figs. 24 and 25. In this sense, the true cause of a crash could be termed a systemic instability. This leads to the possibility that the market anticipates the crash in a subtle self-organized and cooperative fashion. Our theory of collective behavior predicts robust signatures of speculative phases of financial markets, both in accelerating bubbles and decreasing prices. These precursory patterns have been documented for essentially all crashes on developed as well as emergent stock markets (see Sornette [64] for an exhaustive review). Accordingly, the crash of October 1987 is not unique but a representative of an important class of market behavior, underlying also the crash of October 1929 [65] and many others [66,64].

The development of a given financial bubble releases precursory "fingerprints" observable in the stock market prices [32,64]. These fingerprints have been modeled by "log-periodic power laws" (LPPL), which are mathematical patterns associated with the mathematical generalization of the notion of fractals to complex imaginary dimensions [67]. We refer to the book of Sornette [64] for a detailed description and the review of many empirical tests and of several forward predictions. In particular, Johansen and Sornette predicted in January 1999 that Japan's Nikkei index would rise 50 percent by the end of that year, at a time when other economic forecasters expected the Nikkei to continue to fall, and when Japan's economic indicators were declining. The Nikkei rose more than 49 percent during that time. Johansen and Sornette also successfully predicted several short-term changes of trends in the US market and in the Nikkei.

Figs.26-29 show four predictions that were recently issued by our group at ETH Zurich. The first three have ended and were successful. At the time of writing, the last one on the Shanghai market is still running.

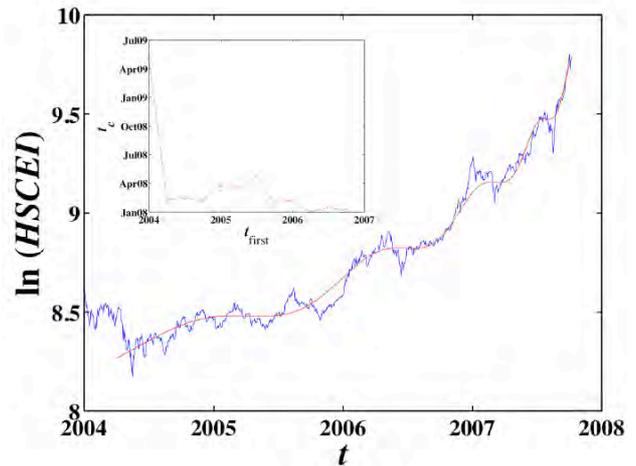

Fig 26 (September 2007) Analysis with the LPPL model of the Hang Seng China Enterprises Index (HSCEI) which led to (i) a diagnostic of an on-going bubble and (ii) the prediction of the end of the bubble in early 2008. I communicated this prediction on 19 October 2007 at a prominent hedge-fund conference in Stockholm. The participants, all supposedly savvy investors and managers, told me that this was impossible because, in their opinion, the Chinese government would prevent any turmoil on the Chinese stock market until at least the end of the Olympic Games in Beijing (August 2008). About a month after my presentation, the HSCEI fell by 20% and the subsequent six months led to a depreciation of more than 65%. The inset shows the predicted time of the end of the bubble as a function of the position of the left-side of the running window over which the analysis is performed. Unpublished work performed with Prof. W.-X. Zhou.

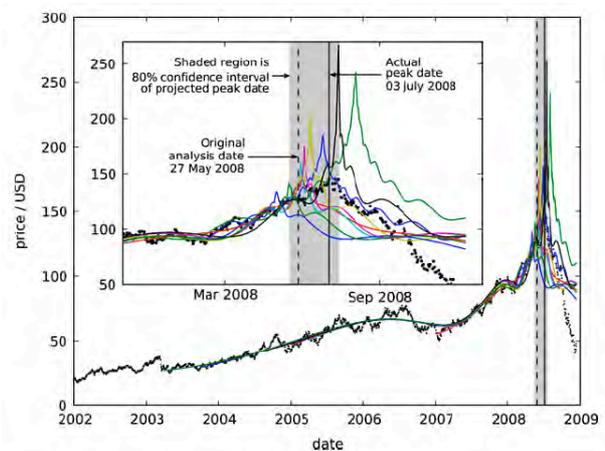

Fig 27 (May 2008) Time series of observed prices in USD of "NYMEX Light Sweet Crude, Contract 1" from the Energy Information Administration of the US Government (see http://www.eia.doe.gov/emeu/international/Crude2.xls) and simple LPPL fits (see text). The oil price time series was scanned in multiple windows. Also shown are dates of our original analysis in June 2008 and the actual observed peak oil price on 3 July 2008. Reproduced from Sornette et al. [68].

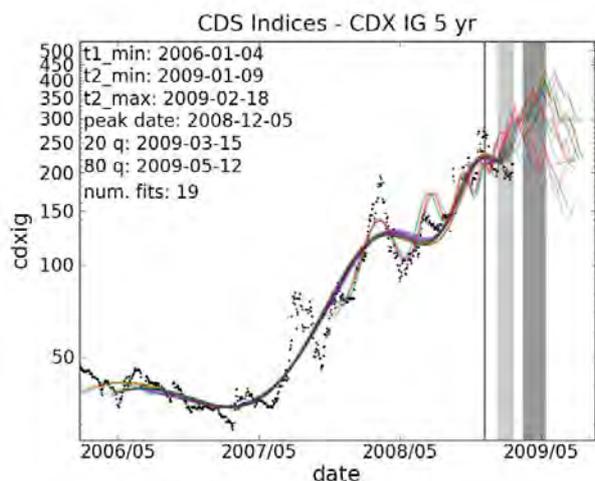

Fig 28 (19 February 2009) Our analysis has been performed on data kindly provided by Amjed Younis of Fortis Bank on 19 February 2009. It consists of 3 data sets: credit default swaps (CDS); German bond futures prices; and spread evolution of several key euro zone sovereigns. The date range of the data is between 4 January 2006 and 18 February 2009. Our log-periodic power law (LPPL) analysis shows that credit default swaps appear bubbly, with a projected crash window of March-May, depending on the index used. German bond futures and European sovereign spreads do not appear bubbly. See report at http://www.er.ethz.ch/fco/CDS for more information.

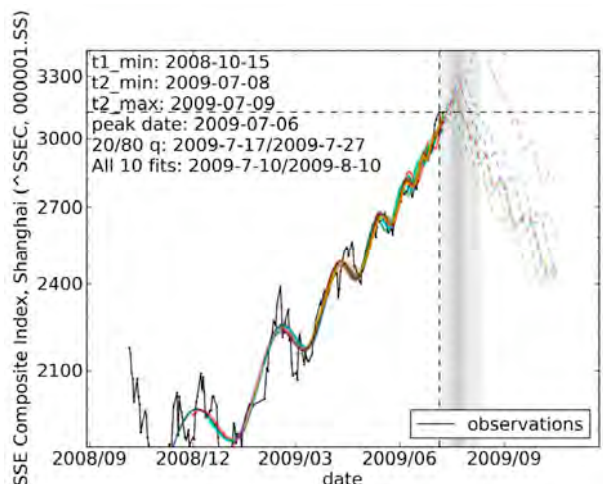

Fig.29 (10 July 2009) Amid the current financial crisis, there has been one equity index beating all others: the Shanghai Composite. Our analysis of this main Chinese equity index shows clear signatures of a bubble build up and we go on to predict its most likely crash date: July 17-27, 2009 (20%/80% quantile confidence interval). Reproduced from Bastiaensen et al. [69].

It is important to stress that our methodology allows us to predict the end of bubbles, but not the crashes per se [33-35]. It is often the case that a bubble bursts into a crash but this is not always the case. The end of a bubble may be a plateau or a slow decay.

Of course, we are not able to predict stock markets with anything close to 100 percent accuracy - just as weather forecasting cannot say with absolute certainty what the weekend weather will be - but our predictions will become more accurate as we refine our methods, as presently undergoing with the new Financial Crisis Observatory [70]. The Financial Crisis Observatory is a scientific platform aimed at testing and quantifying rigorously, in a systematic way and on a large scale the hypothesis that financial markets exhibit a degree of inefficiency and a potential for predictability, especially during regimes when bubbles develop.

Stock market crashes are often unforeseen by most people, especially economists. One reason why predicting complex systems is difficult is that we have to look at the forest rather than the trees, and almost nobody does that. Our approach tries to avoid that trap. From the tulip mania, where tulips worth tens of thousands of dollars in present U.S. dollars became worthless a few months later, to the U.S. bubble in 2000, the same patterns occur over the centuries. Today we have electronic commerce, but fear and greed remain the same. Humans remain endowed with basically the same qualities today as they were in the 17th century.

Our methodology provides an original framework to analyze the origin and consequences of the financial crisis of 2008 [71], which started with an initially well-defined epicenter focused on mortgage backed securities (MBS), and which has been cascading into a global economic recession, whose increasing severity and uncertain duration has led and is continuing to lead to massive losses and damage for billions of people. Sornette and Woodard [71] have presented evidence and have articulated a general framework that allows one to diagnose the fundamental cause of the unfolding financial and economic crisis: the accumulation of several bubbles and their interplay and mutual reinforcement has led to an illusion of a "perpetual money machine" allowing financial institutions to extract wealth from an unsustainable artificial process. Taking stock of this diagnostic, Sornette and Woodard [71] found that many of the interventions to address the so-called liquidity crisis and to encourage more consumption are ill-advised and even dangerous, given that precautionary reserves were not accumulated in the "good times" but that huge liabilities were. The most "interesting" present times constitute unique opportunities but also great challenges, for which a few recommendations can be offered.

Bubbles and crashes are ubiquitous to human activity: as humans, we are rarely satisfied with the Status Quo; we tend to be over-optimistic with respect to future prospects and, as social animals, we herd to find comfort in being (right or wrong) with the crowd. This leads to human activities being punctuated by bubbles and their corrections. The bubbles may come as a result of expectations of the future returns from new technology, such as in the exploration of the solar system, of the human biology or new computer and information technologies. I contend that this trait allows us as a species to take risks to innovate with extraordinary successes, which would not arise otherwise [72,73]. Thus, bubbles and crashes, the hallmark of humans, are perhaps our most constructive collective process. But they may also undermine our quest for stability. We thus have to be prepared and adapted to the systemic instabilities that are part of us, part of our collective organization, ... and which will no doubt recur again perhaps with even more violent effects in the coming decade.

## 5. Conclusions

We have presented supporting evidence for the concept that meaningful outliers (called "dragon-kings") coexist with power laws in the distributions of event sizes under a broad range of conditions in a large variety of systems. These dragon-kings reveal the existence of mechanisms of self-organization that are not apparent otherwise from the distribution of their smaller siblings.

This leads to two consequences, one pessimistic and the other one more optimistic. The first one is the unavoidable evidence that extreme events occur much more often than would be predicted or expected from the observations of small, medium and even large events. Thus, catastrophes and crises are with us all the time. On the other hand, we have argued that the dragon-kings reveal the presence of special mechanisms. These processes provide clues that allow us to diagnose the maturation of a system towards a crisis, as we have documented in a series of examples in various systems.

We have emphasized the use of the concept of a "phase transition – bifurcation – catastrophe – tipping – point," which is crucial to learn how to diagnose in advance the symptoms of the next great crisis, as most crises occur under only smooth changes of some control variables, without the need for an external shock of large magnitude.

The validation of the ideas proposed here is on-going with the creation of the Financial Crisis Observatory [70] using the method of Sornette et al. [74,74].

**Acknowledgements**: This text summarizes a conceptual framework that has matured over the years, while I was working on the characterization and prediction of crises in a variety of systems. This synthesis owes much to my collaborators, whose names appear in the cited papers of the reference lists, especially Anders Johansen from 1995 to 2002, Wei-Xing Zhou since 2002 and more recently Ryan Woodard at ETH Zurich. This is an opportunity to express my extreme gratitude to them. Many of the graphs shown here and the associated discussions have been presented in various seminars and lectures that I gave on various occasions in the last few years. I acknowledge financial support from the ETH Competence Center 'Coping with Crises in Complex Socio-Economic Systems' (CCSS) through ETH Research Grant CH1-01-08-2.